\documentclass[12pt]{article}

\usepackage{amsmath}
\usepackage{caption}
\usepackage{subfig}
\usepackage{graphicx} 
\usepackage{amsmath}
\usepackage{graphics} 
\usepackage{anysize}
\usepackage{multirow}
\usepackage{wrapfig}
\usepackage{float}
\usepackage[usenames]{color}
\usepackage[hidelinks]{hyperref}

\usepackage{scicite}
\usepackage{times}
\usepackage[title]{appendix}

\newenvironment{sciabstract}{%
\begin{quote} \bf}
{\end{quote}}

\newcounter{lastnote}

\title{Atomic Force Microscopy Simulations for CO-functionalized tips with Deep Learning} 

\author
{Jaime Carracedo-Cosme$^{1,2}$, Rub\'en P\'erez$^{1,3,\ast}$\\
\\
\normalsize{$^{1}$Departamento de F\'isica Te\'orica de la Materia Condensada,}\\ \normalsize{Universidad Aut\'onoma de Madrid, E-28049 Spain}\\
\normalsize{$^{2}$ Strategy Big Data S.L.,}\\
\normalsize{C. de Manuel Tovar, 33, 28034 Madrid, Spain}\\
\normalsize{$^{3}$Condensed Matter Physics Center (IFIMAC),}\\ 
\normalsize{Universidad Aut\'onoma de Madrid, E-28049 Madrid, Spain}\\
\normalsize{$^\ast$ ruben.perez@uam.es}
}

\date{}

\usepackage[acronym]{glossaries}
\newacronym{AFM}{AFM}{Atomic Force Microscopy}
\newacronym{NCAFM}{NC--AFM}{non--contact Atomic Force Microscopy}
\newacronym{HRAFM}{HR--AFM}{High Resolution Atomic Force Microscopy}
\newacronym{AFMD}{QUAM--AFM}{}

\newacronym{VAE}{VAE}{Variational Autoencoder}
\newacronym{CNN}{CNN}{Convolutional Neural Network}
\newacronym{MLP}{MLP}{Multilayer Perceptron}
\newacronym{RNN}{RNN}{Recurrent Neural Network}
\newacronym{NLP}{NLP}{Natural Languaje Procesing}
\newacronym{M-RNN}{M--RNN}{Multimodal Recurrent Neural Network}
\newacronym{LSTM}{LSTM}{Long Short--Term Memory}
\newacronym{GRU}{GRU}{Gated Recurrent Unit}
\newacronym{NN}{NN}{Neural Networks}
\newacronym{GUI}{GUI}{Graphical User Interface}
\newacronym{XOR}{XOR}{exclusive--or}
\newacronym{MARNN}{AM--RNN}{Attribute Multimodal Recurrent Neural Network}
\newacronym{SELFIES}{SELFIES}{Self-Referencing Embedded Strings}
\newacronym{GPU}{GPUs}{Graphical Processing Units}

\newacronym{AvgPool}{AvgPool}{Average Pool Layer}
\newacronym{MaxPool}{MaxPool}{Maximum Pool Layer}

\newacronym{lrelu}{LReLU}{Leaky ReLU}
\newacronym{relu}{ReLU}{Rectified Linear Unit Activation Function}
\newacronym{elu}{ELU}{Exponential Linear Unit}
\newacronym{MSE}{MSE}{Mean Squared Error}
\newacronym{MAE}{MAE}{Mean Absolute Error}
\newacronym{RMSE}{RMSE}{Root Mean Square Error}

\newacronym{RMSProp}{RMSProp}{Resilent Mean Square Propagation}
\newacronym{SGD}{SGD}{Stochastic Gradient Descent}
\newacronym{Adagrad}{Adagrad}{Adaptive Gradient Descent}
\newacronym{Nadam}{Nadam}{Nesterov Adaptive Moment Estimator}
\newacronym{Adam}{Adam}{Adaptive Moment Estimator}
\newacronym{CGAN}{CGAN}{Conditional Generative Adversarial Network}
\newacronym{GAN}{GAN}{Generative Adversarial Network}

\newacronym{lr}{lr}{learning rate}
\newacronym{FC}{FC}{Fully connected}
\newacronym{IDG}{IDG}{Image Data Generator}
\newacronym{PPM}{PPM}{Probe Particle Model}
\newacronym{FDBM}{FDBM}{Full Density Based Model}

\newacronym{IUPAC}{IUPAC}{International Union of Pure and Applied Chemistry}
\newacronym{softmax}{Softmax}{Soft Approximation of Max}
\newacronym{AI}{AI}{Artificial Intelligence}

\newacronym{dAFM}{dAFM}{Dynamic Atomic Force Microscopy}
\newacronym{AM-AFM}{AM--AFM}{Amplitude Modulation Atomic Force Microscopy}
\newacronym{CH}{CH}{Constant Height}
\newacronym{DFT}{DFT}{Density Functional Theory}
\newacronym{ES}{ES}{Electrostatic}
\newacronym{SR}{SR}{Short--Range}
\newacronym{vdW}{vdW}{van der Waals}
\newacronym{PTCDA}{PTCDA}{perylene-tetracarboxylic-dianhydride}
\newacronym{FM-AFM}{FM--AFM}{Frequency Modulation Atomic Force Microscopy}
\newacronym{KPFM}{KPFM}{Kelvin Probe Force Microscopy}

\newacronym{LJ}{LJ}{Lennard--Jones}
\newacronym{NC}{NC}{Non--Contact}
\newacronym{NTCDI}{NTCDI}{Naphthalenetetracarboxylic diimide}
\newacronym{PAW}{PAW}{Projector--Augmented--Wave}
\newacronym{PES}{PES}{Potential Energy Surface}
\newacronym{SAM}{SAM}{Self--Assembled Monolayer}
\newacronym{SPM}{SPM}{Scanning Probe Microscopy}
\newacronym{STM}{STM}{Scanning Tunneling Microscopy}
\newacronym{STS}{STS}{Scanning Tunneling Spectroscopy}
\newacronym{SWCNT}{SWCNT}{Single Wall Carbon Nanotubes}
\newacronym{UHV}{UHV}{Ultra--High Vacuum}
\newacronym{LT}{LT}{Low Temperature}
\newacronym{VASP}{VASP}{Vienna Ab initio Simulation Package}
\newacronym{XC}{XC}{Exchange and Correlation}
\newacronym{HR}{HR}{High--Resolution}
\newacronym{TH}{TH}{Tersoff--Hamann}
\newacronym{OpenMX}{OpenMX}{Open source package for Material eXplorer}
\newacronym{C60}{C$_{60}$}{Buckminsterfullerene}
\newacronym{CVD}{CVD}{Chemical Vapor Deposition}
\newacronym{bfA}{bfA}{Breitfussin A}
\usepackage{gensymb}

\newacronym{BLEU}{BLEU}{Bilingual Evaluation Understudy}
\newacronym{M-RNN-AT}{M--RNN$_{A}$}{Multimodal Recurrent Neural Network predicting attributes}
\newacronym{mDBPc}{mDBPc}{meso--Dibenzoporphycene}

\usepackage{color}
\usepackage[nameinlink]{cleveref}

\begin{document} 


\baselineskip24pt

\maketitle 

\begin{sciabstract}

Atomic Force Microscopy (AFM) operating in the frequency modulation mode with a metal tip functionalized with a CO molecule images the internal structure of molecules with an unprecedented resolution.  The interpretation of these images is often difficult, making the support of theoretical simulations important. Current simulation methods, particularly the most accurate ones, require expertise and resources to perform ab initio calculations for the necessary inputs (i.e charge density and electrostatic potential of the molecule). 
Here, we propose an efficient and simple alternative to simulate these AFM images based on a \gls{CGAN}, that avoids all force calculations, and uses as the only input a 2D ball--and--stick depiction of the molecule.  We discuss the performance of the model when optimized using different training subsets. Our  \gls{CGAN} reproduces accurately the intramolecular contrast observed in the simulated  images for quasi--planar molecules, but has significant limitations for molecules with a significant internal torsion, due to the strictly 2D character of the input.

\end{sciabstract}

\newpage

\section{Introduction}


%
\gls{AFM}~\cite{BinnigPRL1986} operating in one of the dynamic modes~\cite{PerezSurfSciRep2002,GiessiblRevModPhys2003} is one of the most powerful techniques for imaging and manipulating materials at the nanoscale.  Combining the atomic resolution provided by the frequency modulation mode with the functionalization of the metal tip with a CO molecule, FM-AFM is able to produce a striking view of  the internal structure of molecules~\cite{GrossScience2009}.  We refer to this combination as \gls{HRAFM}. This development, together with the AFM ability to address individual molecules~\cite{GrossScience2009,PavlicekNatRev2017}, has opened up new research avenues, 
including  the individual discrimination of hundreds of molecules in complex mixtures~\cite{SchulerJACS2015}, the determination of the bond order for each of the bonds in large polycyclic aromatic hydrocarbons~\cite{GrossScience2012},  the visualization of atomic-scale charge distributions~\cite{GrossScience2009b,MohnNatNano2012}, and the characterization  of the reaction process and the intermediate products in on-surface chemical reactions~\cite{deOteyzaScience2013,KawaiNatComm2016b,kawai2017competing,schulz2017precursor}.


%
Understanding the contrast in AFM images is a big challenge. 
The frequency shift  \(\Delta f\) measured in the experiments depends on the complex interplay between the total force acting on the tip during the whole oscillation cycle~\cite{GiessiblPRB1997} and different  operational parameters such as the tip-sample distance,  the oscillation amplitude of the cantilever, the cantilever stiffness, and on the details of the attachment of  the CO molecule to the metal tip, that result in different values of the CO tilting stiffness.  
A combination of experiments and theoretical simulations have paved the way for the identification of key factors controlling  \(\Delta f\). Earlier work showed that the CO-terminated tip \gls{AFM} contrast is mainly controlled by the Pauli repulsion between the lone pair of the oxygen atom in the CO molecule and the charge density of the sample~\cite{GrossScience2009,Moll2010}. This main contribution is modulated by the interaction with the sample's electrostatic field~\cite{VanDerLitPRL2016,HapalaNatComm2016,EllnerNanoLett2016}, and the effect of both force components is enhanced by the probe tilting~\cite{GrossScience2012,HapalaPRB2014,ellner2019molecular}. 
The complex interplay of these interactions, particularly in the case of molecular systems --where they depend on the structure, chemical composition and internal torsion-- makes the interpretation of the experimental features highly non trivial. AFM simulation models with different complexity and accuracy~\cite{MollNJP2012, HapalaPRB2014,HapalaPRL2014,GuoJPCC2015,SakaiNanoLett2016,EllnerNanoLett2016,LeePRB2017,ellner2019molecular} have been developed to compute theoretical \gls{AFM} images using as input the geometry of the molecule. They have allowed to fully understand not only the intramolecular contrast~\cite{MollNJP2012, HapalaPRB2014,HapalaPRL2014,GuoJPCC2015,SakaiNanoLett2016} but also the imaging of intermolecular features on hydrogen and halogen bonded systems~\cite{ellner2019molecular,Tschakert2020natcomm,zahl2021TMA}. 
%

%
While some of these simulation methods are extremely fast, the most accurate ones, retaining a precision similar to density functional theory (DFT) in the determination of the tip--sample forces, require at least a calculation of the charge densities of the tip and sample and the electrostatic potential of the sample. This can be done with standard DFT codes but it's time consuming and requires some theoretical skills that are not always available within experimental groups. 
%
%
Data bases of theoretical AFM images calculated with accurate simulation methods are an alternative for the interpretation of experimental images. We have recently introduced
QUAM--AFM~\cite{QUAM-AFM_repository,QUAM-AFM_repository_CM}, a data set generated from 686K organic molecules, whose geometries have been downloaded form the \href{https://pubchem.ncbi.nlm.nih.gov}{PubChem} repository~\cite{bolton1,kim2016pubchem}. The selected molecules contain the four basic elements of organic chemistry (carbon, hydrogen, nitrogen and oxygen) plus some other less common elements which are still frequent on organic compounds like sulfur, phosphorus and the halogen atoms (fluorine, chlorine, bromine and iodine), and include the most relevant structures and chemical moieties.  In order to support a broad range of experimental conditions, QUAM--AFM contains 165M AFM images simulated for each of those molecules using 240 different combinations of operational parameters (tip-sample distance, oscillation amplitude, and CO tilting stiffness). They represent a computational effort that exceeds 2.5 million hours. 
Given the comprehensive variety of molecular structures provided by QUAM--AFM, it is likely that AFM images for a molecule similar to the one considered in a hypothetical experiment are available, but identifying this related compound would not be easy, even with the search options provided within the data set. Furthermore, it is impossible to account for the factorial growth of possible structures compatible with the combination rules of organic chemistry.   

%
In this paper, we introduce a new, extremely fast, and easy--to--use approach to simulate \gls{AFM} images based on a deep learning model, that avoids all force calculations, and uses as the only input a 2D ball--and--stick depiction of the molecule, where balls of different colors and radii represent the atomic species, and sticks correspond to the covalent bonds.
In particular, we have used a \gls{CGAN}~\cite{isola2017image}, a neural network model that maps each input image to an output image. 
Our \gls{CGAN} has been designed and trained to take as input the  2D ball--and--stick image of a molecule and to produce as output a set of 10 constant--height \gls{AFM} images at different tip-sample distances. As explained below, training, validation and testing has been carried out using information from QUAM--AFM, which provides both the ball-and-stick depiction and 3D stacks of constant-height \gls{HRAFM} images simulated under different operation conditions for each of the molecules. The analysis of the results has been performed by visual comparison of the target images (from QUAM--AFM) for molecules not included in the training set and those generated by our \gls{CGAN}.  It clearly demonstrates that the model provides an efficient and simple alternative to simulate  \gls{HRAFM} images for a relevant range of tip--sample distances, identifies the best data for the training,  and highlights its limitations for molecules with a significant internal torsion, due to the strictly 2D character of the input.

The \gls{CGAN} has been used in a wide variety of applications, ranging from the medical field with the detection of covid-19~\cite{loey2020deep} or brain tumors~\cite{yu2018} from the results of different medical imaging techniques, through common deep learning fields such as synthetic data generation~\cite{Torkzadehmahani2019,ramponi2018t}, image denoising~\cite{Ma2018} or person re-identification~\cite{tang2020cgan}.  
Deep learning has been already used in the AFM field.  \gls{CNN} has been employed for the determination of molecular geometries~\cite{alldritt2020automated}  and the prediction of electrostatic fields~\cite{Foster2022ACSNano}  from \gls{HRAFM} images, while graph neural networks (GNNs) have been applied to extract molecular graphs~\cite{Foster2022MRSbulletin}. The combination of AFM imaging with Bayesian Inference and DFT calculations has been used  to determine the adsorption configurations for a known molecule~\cite{jarvi2021}. In previous work, we have demonstrated that it is possible to achieve a complete chemical identification of the structure and composition of a molecule from a 3D stack of constant--height  \gls{HRAFM} images using two different approaches: (i) a \gls{M-RNN}, that produces as output the IUPAC name of the molecule~\cite{IUPAC_Identification}, and, (ii) a  \gls{CGAN}, that provides a 2D ball--and-stick model of the imaged molecule~\cite{CGAN_Identification}. To the best of our knowledge, deep learning has not been used before to perform theoretical AFM simulations.

The rest of the paper is organized as follows. \Cref{Sec:Methods} describes the structure and operation of the \gls{CGAN} (\cref{Sec:CGAN}) and the training details (\cref{Sec:training}). Results are presented in~\cref{Sec:results}, where the performance of  our \gls{CGAN} is analyzed for different training subsets that are characterized by the maximum internal corrugation of the molecules considered during the training, and by the value of the cantilever oscillation amplitude used for the simulation of the training images.
We end up with the conclusions 
and an outlook for possible future work.

\section{Methods}\label{Sec:Methods}

\subsection{Our CGAN for AFM simulations}\label{Sec:CGAN}

\begin{figure}[p]
\centering
\includegraphics[clip=true, width=0.8\columnwidth]{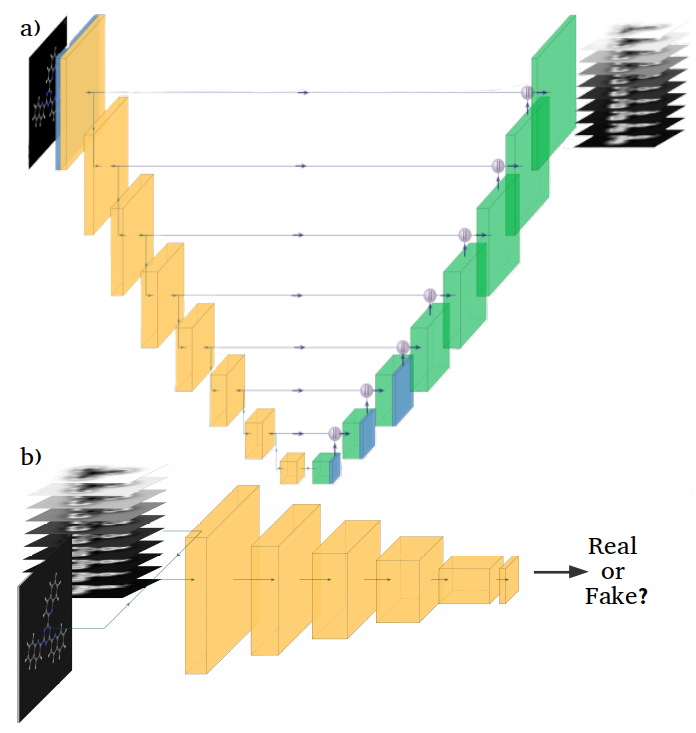}
 \caption{Representation of our \gls{CGAN}, composed of two networks: (a) generator -- with encoder and decoder-- and (b) discriminator.  During training, each ball-and-stick molecular depiction feeds the encoder, which transforms the input into a compressed representation of the molecule to be subsequently reconstructed by the decoder as a stack of 10 constant--height \gls{HRAFM} images for different tip-sample distances. Once the AFM images are generated, the discriminator network tests how good is the simulation produced by the generator network, trying to guess if the if the \gls{HRAFM} images provided as input ( together with the ball-and-stick depiction) are the real ones or have been produced by the generator network. The confrontation of the two networks in a zero-sum game, with the generator improving its performance to fool the discriminator, provides an extremely efficient training. Once training is finished, the discriminator is discarded and the generator is ready to be used to predict the \gls{HRAFM} images for any molecule from its ball-and-stick depiction. 
Regarding structure, yellow boxes in the generator encoder and the discriminator represent blocks consisting of a 2D convolutional layer with batch normalization activated with \gls{lrelu}, the blue ones correspond to dropout layers, while the green boxes in the generator decodes stand for  blocks with a 2D transposed convolutional layer with batch normalization activated with \gls{relu} activation (except for the one previous to the output, that is activated with a hyperbolic tangent function). A detailed description of each layer can be found in appendix~\ref{app:CGAN}.}
\label{Fig:CAN_simulations}
\end{figure}

A \gls{CGAN} includes two sub-networks, known as generator and discriminator (see \cref{Fig:CAN_simulations}). Before describing the detailed block structure of the two networks, we focus on explaining how they operate during the training process. The ball-and-stick depictions and 3D stacks of \gls{HRAFM} images contained in a subset of QUAM--AFM  are used as input and target images during the training (see~\cref{Sec:training} for details on  training hyper parameters).
 
The generator has a U-Net structure, where both the encoder and decoder are convolutional networks (see \cref{Fig:CAN_simulations}). 
In our implementation, each ball-and-stick molecular depiction feeds the encoder, which transforms the input into a compressed representation of the molecule to be subsequently reconstructed by the decoder as a stack of 10 constant--height \gls{HRAFM} images for different tip-sample distances. Once the AFM images are generated, the discriminator network tests how good is the simulation produced by the generator network. 
Either the generator prediction or the real stack of simulated \gls{HRAFM} images, together with the ball-and-stick depiction, are alternatively used to feed the discriminator, which compares patches of the two inputs (the 3D stack and the molecular depiction) to find out if the \gls{HRAFM} images are the real ones or have been produced by the generator network.
The overall loss function of the \gls{CGAN}~\cite{isola2017image}, that quantifies its performance during the training, simultaneously depends on the predictions performed by the generator (comparing the predicted and real images using an L1 norm) and by the discriminator, determining if the images are the predicted or real ones.  In this way, both generator and discriminator try to minimize their losses in a confrontation in which the success of one network forces the other one to improve its predictions. This zero-sum game is a key factor in the performance of  \gls{CGAN}s.
 Once the training is completed, the discriminator is discarded and the generator is ready to be used to predict the \gls{HRAFM} images for any molecule from its ball-and-stick depiction. 

The block structure of  the generator and discriminator is also displayed in \cref{Fig:CAN_simulations}.  Both the generator encoder and discriminator blocks are comprised of a convolutional layer, sequentially followed by a batch normalization and a \gls{lrelu} activation. The generator decoder blocks follow a similar scheme, whereby convolution is replaced by transposed convolution and \gls{relu} is used as the activation function, except in the last block, that is activated with a hyperbolic tangent function. Note that each encoder output feeds the next encoder block and, in addition, the decoder block with the same image size (see \cref{Fig:CAN_simulations}). A detailed description of each of the blocks for the two networks can be found in appendix~\ref{app:CGAN}.

\subsection{CGAN training}\label{Sec:training}



As already mentioned above,  we rely on information from the QUAM--AFM data set for the training of our  \gls{CGAN}. In particular,  we use the ball-and-stick depictions contained in QUAM--AFM to feed the generator and the stack of 10 \gls{AFM} images at different tip-to-sample distances as the target output to compare with the generator predictions. A random value in the range of [0.85,1.15] is selected to apply the zoom to each input-output pair during the training.

\gls{HRAFM} images depend significantly on the internal torsion of the molecule (i.e the differences in height among its constituent atoms), and on operational \gls{AFM} parameters,such as the cantilever oscillation amplitude,  that can be controlled at will during experiment. In our work, we have explored the influence of both factors in the performance of our model. To this end, we have chosen for the training subsets of QUAM--AFM that only include images from molecules with maximum value in the internal height differences in a given range and where a particular value for the oscillation amplitude has been used for the image simulation. Other factors, such as  the torsional stiffness of the CO molecule, that depends on the detailed attachment of the molecule to the metal tip, have been kept fixed (in this case, with a value of 0.4~N/m). 

The different subsets used for the training are described together with the performance of the resulting models in~\cref{Sec:results}. 
All the trainings were performed with the same hyperparameter selection (epochs, batches and batch size) and were optimized in the same way. In particular, a L1 norm was used for the generator --compiled with \gls{MAE} (using the parameter $\lambda=100$ defined by Isola~\textit{et al.}~\cite{isola2017image})--, while the binary cross entropy was used as loss function for the discriminator. The model was minimized by applying batches of 32 inputs with the \gls{Adam} optimiser, where the learning rate and first moment parameters were set to $2\cdot 10^{-4}$ and $0.5$ respectively. The model was trained during 100K iterations, displaying 300 predictions of the validation set to estimate the optimal weights every 10.000 iterations.

\section{Results and Discussion}\label{Sec:results}

As mentioned above, the performance of  our \gls{CGAN} has been analyzed for different training subsets. Each subset only includes  images for  molecules with internal corrugation below a given value and that have been simulated with a certain value of the cantilever oscillation amplitude. 
Once trained, we have tested the model with a total of 1.000 structures that were randomly selected from the test set. 
The test was conducted by comparing predictions from the trained generator network with the ground truth (the real simulated \gls{HRAFM} images) visually.
%

\begin{figure}[t!]
\centering
\includegraphics[clip=true, width=1.0\columnwidth]{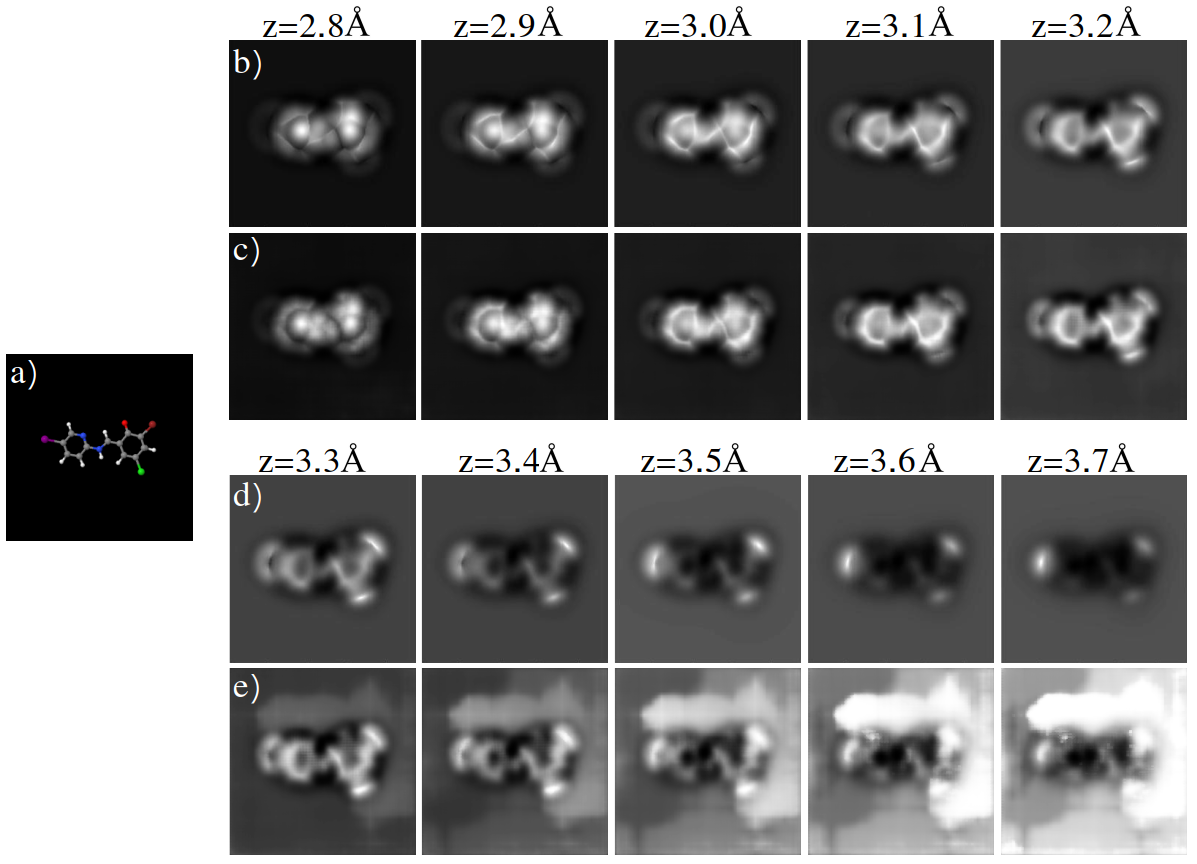}
\caption{(a) Ball--and-stick depiction, (b) and (d) \gls{HRAFM} simulations, and (c) and (e) \gls{CGAN} predictions for 2-bromo-4-chloro-6-[[(5-iodopyridin-2-yl)amino]methylidene]cyclohexa-2,4-dien-1-one. The chemical species represented by balls in (a) are carbon (grey), hydrogen (white), bromine (maroon), chlorine (lime), iodine (purple), nitrogen (blue) and oxygen (red). The \gls{HRAFM} images have been calculated with an oscillation amplitude of 0.4 \r{A} and a tilting stiffness of 0.4 $N/m$. The same simulation parameters were used to generate the images with which the model was trained. Only molecules whose maximum height difference between atoms is less than 0.5 \r{A} were included in the training set.}
\label{Fig:Test_Background}
\end{figure}

\Cref{Fig:Test_Background}  shows the stack of  \gls{HRAFM} images predicted for 2-bromo-4-chloro-6-[[(5-iodopyridin-2-yl)amino]methylidene]cyclohexa-2,4-dien-1-one by the model trained with images simulated with an oscillation amplitude of 0.4 \r{A} and a tilting stiffness of 0.4~$N/m$ for molecules with corrugation --the maximum height difference between the constituent atoms-- below 0.5\r{A}. While the predictions  for the most relevant  tip--sample distances (2.8--3.2 \r{A}, (\cref{Fig:Test_Background}c) are in striking agreement with the real simulations (\cref{Fig:Test_Background}b), at longer distances, they still reproduce reasonably well the intramolecular contrast but fail to display correctly the surrounding background (see \cref{Fig:Test_Background}e and d). 
Tip--sample distances are defined from the oxygen of the CO molecule to the average molecular plane (the average height of the atoms in the molecule), following the prescription used in the QUAM--AFM data set used for the training.
Starting with the intramolecular contrast, the model predicts the structure of the \gls{AFM} images with high accuracy. The molecule shown in \cref{Fig:Test_Background} contains C, H, N, O, Cl, Br and I, almost all the chemical species in the QUAM-AFM dataset. For each of the halogens, the model correctly predicts the characteristic image feature associated with these atoms~\cite{ellner2019molecular,Tschakert2020natcomm}, a rather elongated oval  perpendicular to the C--halogen bond centered at the position of the halogen, and its evolution with the tip--sample distance. 
Similarly, predictions of the \gls{AFM} features caused by nitrogen (e.g. the sharper vertex induced in the ring by the N lone pair~\cite{ellner2019molecular}) and by the oxygen electronegativity~\cite{zahl2021TMA} are accurately reproduced. 

Continuing the analysis with the contrast shown in the background, we assign the error shown in \cref{Fig:Test_Background}d for large tip-sample distances to two factors: 
Firstly, \gls{CGAN} introduces a \gls{MAE} (or L1 norm as denoted in ref.~\cite{isola2017image}) in the loss function. During training, the optimisation process seeks a balance between the sharpness of the output image and the objects in it. The generator must not only fool the discriminator, but also approach the ground truth output in a L1 sense. A priori the results are better with this combination. However, by using the L1 norm in the minimisation, we also introduce a blur filter, well known in deep learning \cite{pmlr-v48-larsen16,isola2017image}. This contribution is responsible for the white patches on the grey background. 
Secondly, the \gls{AFM} contrast depends on both the chemical species in the sample and their distance to the CO molecule at the tip. Since, during the training of this model, we have fed it with ball-and-stick depictions and \gls{HRAFM} images of molecules with a corrugation below 0.5\r{A} height difference, the range of distances explored to differentiate the moleculeformx its surroundings is limited. We believe, that with this training,  the model does not have enough information to predict the contrast change in the background accurately.

\begin{figure}[t!]
\centering
\includegraphics[clip=true, width=1.0\columnwidth]{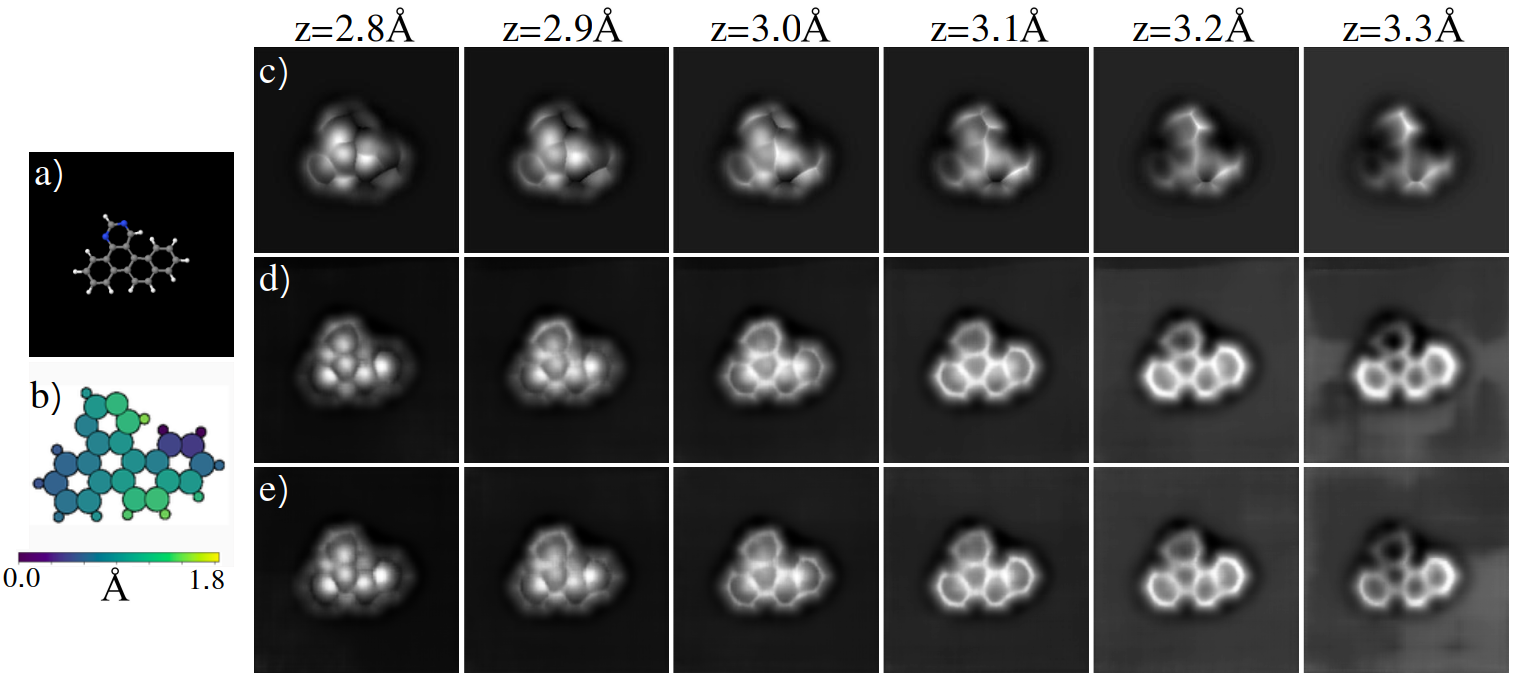}
\caption{ (a)  Ball--and-stick depiction, and (b) height map for the 9,11-diazapentacyclo[12.8.0.0\textsuperscript{2,7}.0\textsuperscript{8,13}.0\textsuperscript{15,20}]docosa-1(14),2,4,6,8,10,12,15,17,19,21-undecaene molecule.  
(c) \gls{HRAFM} simulations, (d) and (e) \gls{CGAN} predictions for models trained with molecules with a maximum height difference between atoms of 0.5 and 1.1 \r{A} respectively. Simulation parameters for (c) and for the images used in the training are 0.4 $N/m$ for the tilting stiffness and 0.4 \r{A} for the oscillation amplitude. The color code for the chemical species in (a) is the same one used in \cref{Fig:Test_Background}.}
\label{Fig:Test_Height}
\end{figure}



\begin{figure}[t!]
\centering
\includegraphics[clip=true, width=1.0\columnwidth]{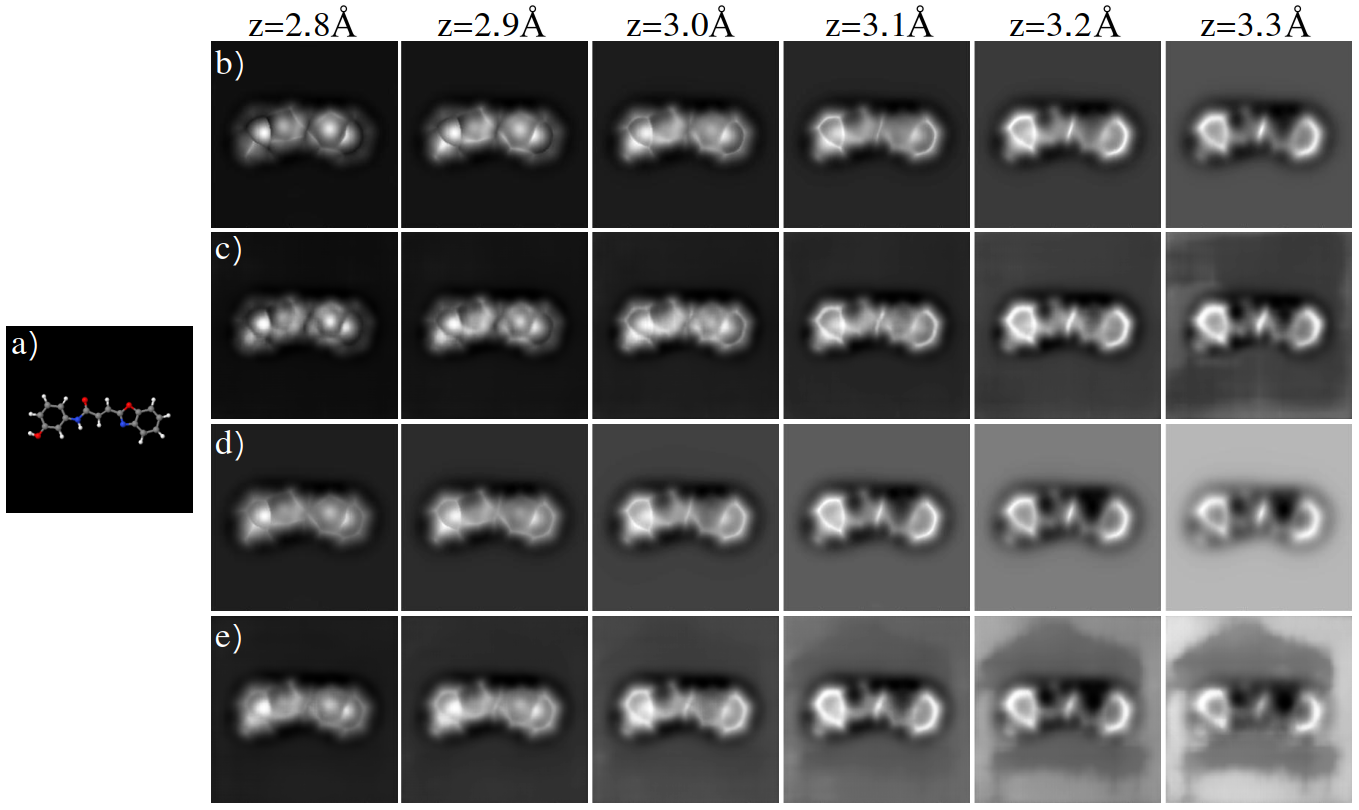}
\caption{(a)  Ball--and-stick depiction for
3-(1,3-benzoxazol-2-yl)-N-(3-hydroxyphenyl)prop-2-enamide, used as input to the model. The color code for the chemical species in (a) is the same one used in \cref{Fig:Test_Background}. (b) \gls{AFM} simulations with amplitude 0.4 \r{A} and  (c) the corresponding prediction, (d) simulations with amplitude 1.0 \r{A} and (e) the prediction. These results show that, while the intramolecular contrast is correct in both cases,  the model improves its accuracy in the prediction of the background when images calculated with smaller oscillation amplitudes are used in the training.}
\label{Fig:Test_Amplitude}
\end{figure}

Considering the limited  information provided by a 2D ball--and--stick depiction for molecules with non-negligible internal torsion, we have explored the \gls{CGAN} performance for these molecules, comparing the results produced by models optimized with two different training subsets:  one includes  molecules with maximum corrugations below 0.5 \r{A} (this is the one also used in~\cref{Fig:Test_Background}, while the other extends the posible internal height difference to  1.1 \r{A} . 
 \Cref{Fig:Test_Height}  shows the predictions of these two models for the 9,11-diazapentacyclo[12.8.0.0\textsuperscript{2,7}.0\textsuperscript{8,13}.0\textsuperscript{15,20}]docosa-
\linebreak 1(14),2,4,6,8,10,12,15,17,19,21-undecaene molecule.
Despite being a molecule with a marked 3D structure, in both cases the \gls{CGAN} is predicting images that would correspond to a planar structure, i.e. in the prediction, the intramolecular bonds do not reflect the torsion of the molecule. However, differences can be seen in the predictions coming from each different training subset. For example, the contours in the closest image (2.8 \r{A}) are better defined when molecules with smaller height differences are used in the training of the model (compare~\cref{Fig:Test_Height}~(d) with \cref{Fig:Test_Height}~(e)). The prediction of contrast change in the overall test is also more accurate when the training is performed with flatter molecules. 
Since the new model is still unable to capture the internal torsion of the molecule and does not improve the description of the contrast evolution, training with molecules with a strong torsion is unfavorable. 
This result can be understood considering how the optimization process works. We are introducing an additional variable in the final result but we are not providing the corresponding input information. Therefore, the generator is unable to reproduce the output. This leads to an increase in the value of the loss function when calculating the L1 norm between the prediction and the ground truth. In turn, the information about the torsion is not provided to the discriminator either, so the generator learns to fool the discriminator very easily. This leads to a minimisation offset as the loss function is not balanced.

Finally, we consider the effect of changing the oscillation amplitude used in the training images. 
%
%
\Cref{Fig:Test_Amplitude} shows the predictions for 3-(1,3-benzoxazol-2-yl)-N-(3-hydroxyphenyl)prop-2-enamide form models trained with images simulated  with oscillation amplitudes 0.4 \r{A} and 1.0 \r{A} (\cref{Fig:Test_Amplitude}c and e, respectively).  These results are compared to the true simulations with the corresponding oscillations amplitudes (\cref{Fig:Test_Amplitude}b and d, respectively).
Although there are small variations in the intramolecular contrast, the main difference between the simulations is in the image background, that changes significantly with the tip--sample distance when the larger oscillation amplitude is used. The predictions of the \gls{HRAFM} images in \cref{Fig:Test_Amplitude}c and e show that, while the intramolecular contrast is accurately reproduced in both cases, the model improves its performance in the prediction of the background when images calculated with smaller oscillation amplitudes are used in the training.

\section{Conclusions}\label{Sec:conclusions}

We have shown that a  \gls{CGAN} provides an efficient and simple alternative to simulate  \gls{HRAFM} images for a relevant range of tip--sample distances, using as the only input, the 2D ball--and--stick depiction of the molecule. We have determined its performance for different training subsets defined in terms of two key parameters: the maximum internal torsion of the molecules include in the set, and the oscillation amplitude used for the simulation of the training images. 
Our  \gls{CGAN} reproduces accurately the intramolecular contrast observed in the simulated  \gls{HRAFM} images for quasi--planar molecules (with a corrugation of less that 0.5\AA). The prediction of the image background deteriorates for larger tip-sample distances, but it can be improved using small oscillation amplitudes for the training. This problem with the background can be traced back to the L1 norm introduced in the optimization of the generator. While this choice is useful for the combined minimization of 
the generator and the discriminator, it also introduces a blur filter that is responsible for the irregular patches observed in the background. 
Due to the limited information provided by the strictly 2D input, predictions for molecules with significant internal corrugation correspond  to planar structures, failing to capture the contrast associated with the atomic height differences.

Looking into the future, the fact that the background prediction improves when images with small amplitudes are used in the training suggests a possible (but rather complex) alternative to improve the results: to parameterize tip-sample forces using machine learning techniques and, later on, to use them to calculate the frequency shift.  This alternative approach should also reduce the problem of blurred filters in the background because doing the integral to calculate the frequency shift would result in a smoother output. Machine-learning force fields have been already developed for different materials and applications (see~\cite{Tkatchenko2021MLFF} for a review), but given the large number of chemical species involved in organic chemistry, we anticipate it would be difficult to achieve the generality provided by our \gls{CGAN}.
The 2D character of the ball--and--stick depiction limits the molecules that can be accurately simulated with this model. The development of a data input providing information about the relative heights of the atoms is a key point to be explored in the next work.

\begin{appendices}
\section{Detailed description of the CGAN blocks}\label{app:CGAN}

We provide here all the details necessary to reproduce our model. The \gls{CGAN} generator has an input of one image with three channels (RGB) and a target output of 10 images with a single channel (grayscale). The ball-and-stick depiction feeds an initial dropout layer with a rate of 0.4, followed by a 2D convolutional layer with 64 kernels, each one with size (4,4) and stride (2,2) and where padding is applied. The layer is activated with \gls{lrelu} function. This first block of layers is followed by a series of seven similar blocks (represented by yellow boxes in~\cref{Fig:CAN_simulations}) consisting of a sequence of a 2D convolution, followed by a batch normalisation and a \gls{lrelu} activation function with $\alpha=0.2$. All kernels of the convolutions have size (4,4) and are applied with a stride of (2,2). The 2D convolutional layers have 128, 256, 512, 512, 512, 512, 512 kernels, taking as reference the processing direction from the one closest to the input to the one closest to the compressed representation space. The outputs of the activations are used both to feed the next block of the encoder and to feed the decoder block of the same size. The generator decoder blocks, represented by green boxes in~\cref{Fig:CAN_simulations},  consist of another series of layers in which a transposed convolution, a batch normalization, a dropout layer (only in the three layers closest to the space of the compressed representation, see~\cref{Fig:CAN_simulations}), a concatenation with the output of the corresponding encoder block, and, finally, a \gls{relu} activation, except for the 
output layer that is activated with an hyperbolic tangent.

The discriminator consists of a sequence of layers, initiated by a concatenation of all input images (note that we can consider the 10 \gls{AFM} images as a single image with 10 channels). It is followed by a 2D convolutional layer with 64 kernels of size (4,4) and stride of (2,2) activated with \gls{lrelu}. Then, it has four blocks consisting of a 2D convolutional layer, a batch normalization and a \gls{lrelu} activation ($\alpha=0.2$). The convolutions have 128, 256, 512 and 512 kernels with size (4,4) and stride (2,2) respectively. The last layer is a 2D convolution with a single kernel of size (4,4) which is activated with the sigmoid function.

\end{appendices}

\section*{Acknowledgments}
Support from the Comunidad de Madrid Industrial Doctorate programme 2017 under reference number IND2017/IND-7793 and from the Spanish Ministry of Science and Innovation (project PID2020-115864RB-I00) is gratefully acknowledged.  R.P. acknowledges support from the Spanish Ministry of Science and Innovation through the ``Mar\'{\i}a de Maeztu'' Programme for Units of Excellence in R\&D (CEX2018-000805-M).  

\newpage
\bibliographystyle{Science}
\bibliography{DL_AFM_4}
\addcontentsline{toc}{chapter}{Bibliography}

\end{document}